# Analytical and numerical studies of the one-dimensional sawtooth chain

Jian-Jun Jiang [a,*], Yong-Jun Liu [b], Fei Tang [c], Cui-Hong Yang [d], Yu-Bo Sheng [e]

[a]*Department of Physics, Sanjiang College, Nanjing 210012, China*

[b]*School of Physics Science and Technology, Yangzhou University, Yangzhou 225002, China*

[c]*Department of Electronic and Information Engineering, Yangzhou Polytechnic Institute, Yangzhou 225127, China*

[d]*Faculty of Mathematics and Physics, Nanjing University of Information Science and Technology, Nanjing 210044, China*

[e]*Institute of Signal Processing Transmission, Nanjing University of Posts and Telecommunications, Nanjing 210003, China*

**Abstract:** By using the analytical coupled cluster method, the numerical exact diagonalization method, and the numerical density matrix renormalization group method, we investigated the properties of the one-dimensional sawtooth chain. The results of the coupled cluster method based on Néel state demonstrate that the ground state is in the quasi-Néel-long-range order state when $\alpha < \alpha_{c_1}$. The translational symmetry of the ground state varies and the ground state evolves from the quasi-Néel-long-range order state to the dimerized state at the critical point $\alpha_{c_1}$. The dimerized state is stable in the intermediate parameter region and vanishes at another critical point $\alpha_{c_2}$. The results drawn from the exact diagonalization show that the precise critical point $\alpha_{c_1}$ and $\alpha_{c_2}$ can be determined by using the spin stiffness fidelity susceptibility and spin gap separately. We compared the results obtained by using the coupled cluster method based on canted state with those obtained based on spiral state, and found that the ground state of the sawtooth chain is in the quasi-canted state if $\alpha > \alpha_{c_2}$. The results of the coupled cluster method and the density matrix renormalization group method both disclose that the type of the quantum phase transition occurring at $\alpha_{c_2}$ belongs to the first-order transition.

## 1. Introduction

Quantum phase transition driven by quantum fluctuation is one of the prime research topics in condensed matter physics. One-dimensional uniform Heisenberg spin chain with frustration plays an important role in understanding quantum phase transition which occurs at zero temperature because such systems display a wide variety of exotic quantum phases. The quantum fluctuation in the low dimensionality is so strong that the classical magnetic long-range order of the spin

*Corresponding author: Department of Physics, San Jiang College, Nanjing 210012, China.

Tel: +86 25 51663485;    Fax: +86 25 51663485

E-mail address: jianjunjiang@126.com (J.-J. Jiang).

chain melts even in the absence of frustration. A well known example is the one-dimensional spin-1/2 $J_1 - J_2$ chain. For $J_2 = 0$, the excitation spectrum of that model is gapless and it possesses quasi-Néel-long-range order, with algebraically decaying spin correlations [1, 2]. At a finite value of the frustration parameter $\alpha = J_2 / J_1 = 0.241$, the model displays a Kosterlitz-Thouless transition from the quasi-Néel state to a two-fold degenerate dimerized state, which breaks the translational lattice symmetry spontaneously [3, 4]. At the Majumdar-Ghosh point ($\alpha = 0.5$), a tensor product of singlet pairs formed by the nearest neighboring spins is the exact ground state of that system [2]. As $\alpha$ increases further, the ground state of the model exhibits interesting incommensurate spiral correlations [5-7]. Another prototypical case that has attracted considerable attention is the sawtooth chain with nearest-neighbor interaction $J$ and second-neighbor interaction between spins in the same sublattice $\alpha J$. For $\alpha = 1$, the sawtooth chain has been studied extensively by various approximate techniques, such as variational method and exact diagonalization (ED) method [8-12]. It has been found that the ground state of the sawtooth chain is also a two-fold degenerate gapped dimerized state and the elementary excitations are of quantum soliton type [10]. Theoretical interest in the sawtooth lattice was enhanced after it was pointed point that the properties of some new synthetic compounds, such as $YCuO_{2.5}$, can be described by that chain [12]. Recently, people have also paid attention to the relation between the localized magnon states which exist in the sawtooth chain and the low-temperature thermodynamics of the chain [13, 14]. As far as we have known, the case $\alpha \neq 1$ has been discussed only in reference [12], although the case $\alpha = 1$ has been intensively investigated. The results of ED in paper [12] show that the elementary excitation spectrum of the sawtooth chain only has a gap in the region $\alpha_{c_1} < \alpha < \alpha_{c_2}$ and the spin gap disappears at the critical points ($\alpha_{c_1} = 0.4874$ and $\alpha_{c_2} = 1.53$). But, due to strong finite size effect, it is very difficult to analyse the properties of the sawtooth chain in the large $\alpha$ parameter region by using ED. So, the effect of frustration $\alpha J$ on the properties of the chain needs to be further discussed.

In the present paper, we use the analytical coupled cluster method (CCM), numerical ED method, and numerical density matrix renormalization group (DMRG) method (if necessary) to study the sawtooth chain. As shown in figure 1, the model Hamiltonian is

$$H = J\sum_{i=1}^{N/2} S_{2i}\cdot(S_{2i-1}+S_{2i+1}) + \alpha J\sum_{i=1}^{N/2} S_{2i}\cdot S_{2i+2} \quad (1)$$

where $S_{2i}$ , $S_{2i-1}$ and $S_{2i+1}$ are spin-1/2 operators, and the nearest neighbor interaction $J$ and the next nearest neighbor interaction $\alpha J$ are both antiferromagnetic. The number of the unit cells is denoted by *N*/2, and then the total number of sites is *N*. For convenience, in what follows we set $J = 1$. Classically, the sawtooth chain has three ordered phases, comprising a Néel, a canted, and a spiral state. That model has Néel order for $\alpha < \alpha_c = 1/2$. It exhibits a second-order transition from a Néel phase to a canted or a spiral phase as shown in figure 1 at $\alpha_c$. The classical canted state and the spiral state have the same energy. The canted angle $\theta_c$ and the spiral angle $\theta_s$ are both equal to $\cos^{-1}(1/2\alpha)$ in the region $\alpha > \alpha_c$.

Since the classical Néel order is absent in the quantum sawtooth chain at $\alpha = 0$, one can judge reasonably that the sawtooth chain is in the quasi-Néel state in the small $\alpha$ parameter region. To investigate whether the sawtooth chain possesses the above two non-linear states, we resort to CCM which is a powerful tool to obtain valid and reliable results of the ground state for frustrated quantum spin systems with the non-linear quantum corrections [7, 15-23]. In references [24, 25], it was shown that CCM can be used to analyse the dimer and plaquette valence-bond phases of quantum spin systems perfectly. Here, we also apply CCM to the study of the dimerized state of the sawtooth chain. And our main aim of the paper is to give a complete phase diagram of the sawtooth chain by using CCM which can provide results in the thermodynamic limit. To check the results of CCM and obtain accurate critical points of the sawtooth chain, we also used numerical ED and DMRG to discuss the properties of the sawtooth chain.

The paper is organized as follows. In the next section, the details of the application of CCM formalism to the sawtooth chain are described. In section 3, the results of CCM, ED and DMRG are presented. A summary is given in the final section.

## 2. The coupled cluster method applied to the sawtooth chain

In recent years, a quite new method called CCM has been very successfully applied to different quantum spin chains [7, 15-34]. The interested reader can obtain the detailed descriptions of the CCM applied to quantum spin systems in papers [26-28]. Here, we only briefly describe the application of CCM to the sawtooth chain. The starting point of a CCM calculation is to choose a model state $|\phi\rangle$ and this is often a classical spin state. So we chose the Néel state $|\downarrow\uparrow\downarrow\uparrow\cdots\rangle$

for small values of the frustration parameter but canted state $|\downarrow\nwarrow\downarrow\nearrow\cdots\rangle$ or the spiral state $|\uparrow\nearrow\rightarrow\searrow\downarrow\swarrow\cdots\rangle$ for large $\alpha$ as the model state. Since the canted angle (the spiral angle) may be affected by quantum fluctuation, we do not choose the classical canted angle (the classical spiral angle), but consider the canted angle (the spiral angle) as a free parameter and determine it by minimization of the CCM ground state energy in the CCM calculation based on canted state (spiral state). Then we perform a rotation of the local axes of the spins such that all spins in the model state align along the negative *z*-axis. After this rotation, the CCM parameterization of the ket and bra ground states of model (1) are given by [27, 28]

$$
\begin{aligned}
|\psi\rangle &= e^{S}|\phi\rangle, \quad S=\sum_{l=1}^{N}\sum_{i_1,i_2,\cdots i_l} S_{i_1,i_2,\cdots i_l}\, s_{i_1}^{+}s_{i_2}^{+}\cdots s_{i_l}^{+} \\
\langle\tilde{\psi}| &= \langle\phi|\tilde{S}e^{-S}, \quad \tilde{S}=\sum_{l=1}^{N}\sum_{i_1,i_2,\cdots i_l} \tilde{S}_{i_1,i_2,\cdots i_l}\, s_{i_1}^{-}s_{i_2}^{-}\cdots s_{i_l}^{-}
\end{aligned}
\tag{2}
$$

The correlation coefficients $S_{i_1,i_2,\cdots i_l}$ and $\tilde{S}_{i_1,i_2,\cdots i_l}$ contained in the operators $S$ and $\tilde{S}$ can be determined by the following CCM equations [27, 28]

$$
\begin{aligned}
&\langle\phi|s_{i_1}^{-}s_{i_2}^{-}\cdots s_{i_l}^{-}e^{-S}He^{S}|\phi\rangle = 0 \\
&\langle\phi|\tilde{S}e^{-S}\left[H, s_{i_1}^{+}s_{i_2}^{+}\cdots s_{i_l}^{+}\right]e^{S}|\phi\rangle = 0
\end{aligned}
\tag{3}
$$

After the correlation coefficients have been obtained, one can use them to calculate the ground state expectation value of some physical observables of the sawtooth chain. For instance, the ground state energy is given by

$$E_g = \langle\phi|e^{-S}He^{S}|\phi\rangle \tag{4}$$

and the magnetic order parameter which is expressed in the local, rotated spin axes can be written as

$$M = -\frac{1}{N}\sum_{i=1}^{N}\langle\tilde{\psi}|s_i^{z}|\psi\rangle \tag{5}$$

Although the CCM formalism is exact if all spin configurations in the *S* correlation operator are considered, it is impossible in practice because the CCM equation systems would be infinite. A big advantage of the CCM compared to some other methods is the possibility to truncate *S* in a very systematic and reasonable way. Here, we use a quite general approximation scheme called LSUB*n* to truncate the expansion of the operator *S* [27, 28]. In the LSUB*n* approximation, only the configurations involving $n$ or fewer correlated spins which span a range of no more than $n$ contiguous lattice sites are retained. The fundamental configurations retained in the LSUB$n$ approximation can be reduced if we choose the collinear Néel state as the model state because the ground state lies in the subspace $S_{tol}^{z} = \sum_{i=1}^{N} S_{i}^{z} = 0$ and the Hamiltonian of equation (1) commutes with $S_{tol}^{z}$. For the canted state or the spiral state, one can not reduce the fundamental configurations because it is not an eigenstate of $S_{tol}^{z}$. Moreover, numerical complexity of the CCM based on the canted state (the spiral state) increases because the determination of the quantum canted angle (spiral angle) requires the iterative minimization of the ground state energy. Therefore, for the Néel model state, we carry out CCM up to the LSUB14 level, whereas for the canted state or the spiral state, we do this only up to the LSUB8 level.

Besides the ground state properties, CCM can also be used to obtain the spin gap of the sawtooth chain if we choose the collinear Néel state as the CCM's model state. To calculate the spin gap, one should firstly obtain the excited-state wave function $\left|\psi_{e}\right\rangle$ which is determined by linearly applying an excitation operator $X^{e}$ to the ket-state wave function (2) and given by [27]

$$\left|\psi_{e}\right\rangle = X^{e} e^{S}\left|\phi\right\rangle, \quad X^{e} = \sum_{l=1}^{N} \sum_{i_1, i_2, \cdots i_l} \chi_{i_1, i_2, \cdots i_l} s_{i_1}^{+} s_{i_2}^{+} \cdots s_{i_l}^{+}, \tag{6}$$

Then the spin gap $\Delta$ is determined by the lowest eigenvalue of the LSUB$n$ eigenvalue equations. Those equations with eigenvalues $\varepsilon_{e}$ and corresponding eigenvectors $\chi_{i_1, i_2, \cdots i_l}^{e}$ are given by

$$\varepsilon_{e} \chi_{i_1, i_2, \cdots i_l}^{e} = \left\langle\phi\right| s_{i_1}^{-} s_{i_2}^{-} \cdots s_{i_l}^{-} e^{-S}\left[H, X^{e}\right] e^{S}\left|\phi\right\rangle = 0, \tag{7}$$

Analogously to the ground state, we also use the LSUB$n$ approximation scheme to truncate the expansion of the operator $X^{e}$. But the fundamental configurations for the excited state differ from those for the ground state because those two states have different quantum numbers.

As the derivation of the coupled equations or the eigenvalue equations for higher orders of approximation is extremely tedious, we have developed our own programme by using Matlab to automate this process according to the method discussed in papers [27, 28]. The Matlab code with double precision was performed in a private computer. To check the accuracy of our code, we compared our CCM results with those given by Dr. Damian Farnell [35], such as the ground state energy, the canted angle, and the spin gap, and found that our results agree with his.

In order to obtain results in the limit $n \to \infty$, the ‘raw’ LSUB*n* results have to be extrapolated. Although there are no exact extrapolation rules, one can perform the extrapolation according to the empirical experience. We use the following well-tested formulas [15, 20, 34]

$$E_g(n)/N = a_0 + a_1\left(\frac{1}{n^2}\right)$$

$$M(n) = b_0 + b_1 n^{-0.5} + b_2 n^{-1.5} \quad (8)$$

$$\Delta(n) = c_0 + c_1 n^{-1} + c_2 n^{-2}$$

for the ground state energy per spin $E_g / N$, the magnetic order parameter *M* and the spin gap $\Delta$.

## 3. Results

We first discuss the properties of the quasi-Néel and dimerized state of the sawtooth chain by using CCM based on Néel state. As the dimerized state breaks the translational lattice symmetry, it is necessary to assume that the two-spin nearest-neighbor ket-state correlation coefficient connecting two sites inside the unit cell is distinct from that connecting two different unit cells. Similar to reference [24], those two coefficients are called $S_2^a$ and $S_2^b$ here. The results of the above two coefficients at the LSUB14 level of approximation are displayed in figure 2. Figure 3 shows the ground state energy per site $e = E_g / N$ obtained from CCM and ED. The energies calculated by ED are extrapolated to the thermodynamic limit by using the following formula with *N*=16, 20, 24, and 28 spins [36]

$$f(\alpha, N) = f(\alpha) + c_1 \frac{\exp(-N/c_2)}{N^p} \quad (9)$$

where *p*=2. As can be seen from figure 2, the full translation symmetry solution, that is $S_2^a = S_2^b$, is the only solution when $\alpha$ is below a critical point $\alpha_{c_1}$. The ground state energy per site $e$ given by CCM’s symmetry solution agrees well with that obtained by ED when $\alpha < \alpha_{c_1}$, as shown in figure 3. Those results indicates that, for the sawtooth chain, the region of quasi-Néel-long-range order extends up to larger value of $\alpha$ compared with the case of the one-dimensional spin-1/2 $J_1 - J_2$ chain [24]. This result is reasonable because only one sublattice is coupled with frustration $J_2$ in the sawtooth chain. When $\alpha$ exceeds the critical point $\alpha_{c_1}$, besides the symmetry solution, a non-symmetry solution characterized by $S_2^a \neq S_2^b$ appears. That solution exists over a finite range of $\alpha$ and it terminates at a large value $\alpha_t$ beyond which the CCM equations based on Néel state have no real solution. For the symmetry solution, the termination phenomenon also occurs at another value of the parameter $\alpha$. From figure 3, one can clearly see that, in the intermediate parameter region, the ground state energy of the non-symmetry solution compares extremely well to that given by ED, while the ground state energy obtained from the CCM symmetry solution deviates from the ED result. For the special case of $\alpha=1$, we find that all ket-state correlation coefficients contained in formula (2) given by non-symmetry solution equal to zero except for $S_2^a$ ($S_2^a=1$), and the ground state energy is $e=-0.375$, which means that the ground state constructed by CCM is the exact dimerized product state as $\alpha=1$. This finding is consistent with the previous research [9, 10]. So far, one can draw a reasonable conclusion that the dimerized state dominates the property of the ground state of the sawtooth chain in the intermediate parameter region. The results of $\alpha_{c_1}$ at different levels of LSUB*n* approximation are displayed in table 1. By using method introduced in reference [37], reference [12] shows that the precise critical point is $\alpha_{c_1}=0.4874$. Thus, the critical point $\alpha_{c_1}$ obtained from CCM is still higher than the above value even under high order LSUB*n* approximation. The similar phenomenon also occurs in the one-dimensional spin-1/2 $J_1-J_2$ chain [24].

It is known that the quasi-Néel state is gapless, while the dimerized state is gapful. Therefore, one can detect the transition between the quasi-Néel state and the dimerized state by using the parameter spin gap $\Delta$ which is defined as follows

$$\Delta = E_1(S_{tol}^z=1)-E_g(S_{tol}^z=0) \tag{10}$$

where $E_1$ and $E_g$ are the energies of the lowest-lying state with $S_{tol}^z=1$ and $S_{tol}^z=0$. Here, we both used ED and CCM under periodic boundary condition (PBC) to calculate the spin gap. For ED, the extrapolation of the data for system sizes of *N*=16, 20, 24 and 28 to the thermodynamic limit is carried out by using formula (9) with *p*=1. For CCM, we used formula (8) to extrapolate the results of LSUB*n* with *n*={8, 10, 12, 14} to the limit $n\to\infty$. To check whether the extrapolated results of ED and CCM are reliable, we compared our results with some known results. For example, the values for $\Delta$ at $\alpha = 1$ given by ED and CCM are 0.205 and 0.198, respectively. They are both closed to the result given in reference [9].The results of the spin gap $\Delta$ are shown in figure 4. As shown in the inset of figure 4, the value of the spin gap $\Delta$ given by ED does not decrease monotonously with the increase of *N* when $\alpha > 1.2$. Thus, at that parameter region, the extrapolated results of $\Delta$ obtained from ED are unreliable and not shown in figure 4. Similar to previous research [25], we found that the results for the extrapolation of LSUB*n* data are not accurate in the region where $\alpha \approx \alpha_{c_1}$. So, the spin gap obtained from CCM is not displayed in figure 4 at that parameter region. The results of ED disclose that, as expected, $\Delta$ is nearly zero for a finite region. It obviously appears when $\alpha > 0.6$. And the gap increases with $\alpha$ in the region $0.6 < \alpha < 1$. At $\alpha = 1$, it reaches a maximum value. Then, it decreases for large $\alpha$. The spin gap calculated by CCM is in good agreement with that of ED in the parameter region $0 \le \alpha \le 1$. However, the value for the spin gap obtained from CCM still increases with the increase of $\alpha$ when $\alpha$ just exceeds 1. And it reaches a peak at $\alpha \approx 1.03$. As a result, the spin gap calculated by CCM deviates from that of ED when $\alpha > 1$. This phenomenon may mean that the model state we chose is poor at that parameter region.

To investigate the behavior of the spin gap $\Delta$ in the large $\alpha$ region accurately, the gap for lengths *N*=16, 20 and 24 is also calculated for open boundary condition (OBC) and extrapolated to $N \to \infty$ using the method introduced in [38]. The results of the spin gap under OBC shown in figure 5 indicate that the spin gap vanishes at another critical point $\alpha_{c_2} = 1.48$. Therefore, the ground state of the sawtooth chain evolves from the dimerized state to a gapless state at that point. As the spin gap has small finite-size effects when $\alpha > \alpha_{c_2}$, the phenomenon of the vanishment of the spin gap nearly appears in finite systems, as shown in figure 5. Our estimate of the value of $\alpha_{c_2}$ agrees well with the one given in reference [12]. On the contrary, in the quasi-Néel state region, finite size effect in the spin gap is large. Although our extrapolated results of the spin gap is very small in the region $\alpha < 0.6$, it is difficult to detect the critical point $\alpha_{c_1}$ by using the spin gap drawn from ED precisely.

Since the critical value $\alpha_{c_1}$ determined by CCM deviates apparently from the precise value given in reference [12] and it is hard to be obtained accurately by using the traditional parameter spin gap, one tool of quantum-information theory, which is the fidelity susceptibility, is also used to detect the transition from the gapless quasi-Néel state to the gapful dimerized state of the sawtooth chain. Owing to latest advances in quantum information science, people have recently found that, the fidelity susceptibility, a basic notion of quantum information science, can be used to identify the quantum phase transition of many spin models [39-44]. As the fidelity susceptibility is a purely quantum information concept, the advantage of using it to detect quantum phase transition is that no a prior identification of the order parameter is required. The spin stiffness fidelity susceptibility $\chi_\rho$ we use here is coupled to the spin stiffness. To obtain $\chi_\rho$ of the sawtooth chain, a twist $\phi$ should be applied at every bond of Hamiltonian (1) [3]. Then, the following Hamiltonian is obtained

$$\begin{aligned} H = J\sum_{i=1}^{N/2}\left[S_{2i}^z\left(S_{2i-1}^z + S_{2i+1}^z\right) + \frac{1}{2}\left(S_{2i}^+S_{2i-1}^-e^{i\phi} + S_{2i}^-S_{2i-1}^+e^{-i\phi} + S_{2i}^+S_{2i+1}^-e^{i\phi} + S_{2i}^-S_{2i+1}^+e^{-i\phi}\right)\right] \\ + \alpha J\sum_{i=1}^{N/2}\left[S_{2i}^zS_{2i+2}^z + \frac{1}{2}\left(S_{2i}^+S_{2i+2}^-e^{i\phi} + S_{2i}^-S_{2i+2}^+e^{-i\phi}\right)\right] \qquad (11) \end{aligned}$$

The spin stiffness fidelity susceptibility $\chi_\rho$ of the sawtooth chain in the limit $\phi \to 0$ is defined as [3]

$$\chi_\rho = \frac{2\left(1 - \left|<\psi_0(\alpha,\phi=0)|\psi_0(\alpha,\phi)>\right|\right)}{\phi^2} \qquad (12)$$

where $\psi_0(\alpha,\phi)$ is the ground state of Hamiltonian (11) and it can be calculated by ED. The twist $\phi$ is taken to be $0.001$ in the present paper. In reference [3], $\chi_\rho$ was used to estimate the critical value at which the ground state of the spin-1/2 $J_1 - J_2$ chain evolves from the quasi-Néel state to the dimerized state successfully. Thus, $\chi_\rho$ may be used to detect the similar transition existing in the sawtooth chain. The spin stiffness fidelity susceptibility $\chi_\rho / N$ is plotted as a function of $\alpha$ for various systems in figure 6. It can be found that there is a valley in $\chi_\rho / N$. The location of the valley moves towards a big value of $\alpha$ with the increase of *N*. Therefore, similar to the spin-1/2 $J_1 - J_2$ chain, we can use the location of the value to obtain the critical point $\alpha_{c_1}$ at which the phase transition between the quasi-Néel state and the dimerized state of the sawtooth chain occurs in the thermodynamic limit. To obtain $\alpha_{c_1}$, one can use the following finite-size scaling theory [45]

$$\left|\alpha_{\min}(N)-\alpha_{c_1}\right| \propto N^{-1/\upsilon} \tag{13}$$

where $\alpha_{c_1}$ is the critical point in the thermodynamic limit and $\upsilon$ is the critical exponent of the correlation length. Figure 7 shows the values of $\alpha_{\min}$ where $\chi_\rho / N$ has its minimum as a function of $N^{-1/\upsilon}$ ($\upsilon = 1$). Through a numerical fit for sites with $12 \le N \le 28$, it is found that $\alpha_{c_1} = 0.4892$. The critical point obtained from the measurement of $\chi_\rho$ is well consistent with the one given in reference [12], in which $\alpha_{c_1} = 0.4874$.

Because the property of the sawtooth chain can not be analysed by using CCM based on Néel state in the large $\alpha$ parameter region, we also apply the spiral state or the canted state displayed in figure 1 as the CCM's model state at that region. As the quantum spiral angle $\theta_s$ ( the quantum canted angle $\theta_c$) may be different from the classical case, we perform CCM calculations for arbitrary $\theta_s$ ($\theta_c$) and then determine the quantum $\theta_s$ ($\theta_c$) by minimizing $e(\theta_s)$ ($e(\theta_c)$) with respect to $\theta_s$ ($\theta_c$) at a given level of LSUB*n* approximation. In the case of LSUB6 approximation, the ground state energy per site of the spiral CCM solution as a function of $\theta_s$ is shown in figure 8. Curves in that figure disclose that the minimum in the energy only occurs at $\theta_s = \pi$ when the parameter $\alpha$ is below a critical point $\alpha_c$. This result indicates that the CCM based on spiral state only has Néel solution if $\alpha < \alpha_c$. For frustrating couplings $\alpha \ge \alpha_c$, apart from the Néel solution, the CCM also has a spiral solution because a second minimum at $\theta_s \ne \pi$ emerges. As $\alpha$ exceeds another critical point $\alpha_{c_2}$, a canted state solution characterized by $\theta_c \ne 0$ also appears and this is shown in figure 9. The similar phenomenon is also observed in other level of LSUB*n* approximation. The CCM results of the ground state energy per spin $e$ based on spiral state or the canted state are shown in figure 10. One can see that, although the spiral solution always gives the lower energy in the case of LSUB4 approximation in the whole parameter region, the curve for the spiral solution and the curve for the canted solution cross at a critical point $\alpha_{cross}$ at the LSUB6 or LSUB8 level of approximation. In the case of LSUB*n* approximation with $n > 4$, the lower energy is determined by the spiral solution when $\alpha < \alpha_{cross}$ and it is given by the canted solution if $\alpha > \alpha_{cross}$. Moreover, the location of $\alpha_{cross}$ moves nearer to the critical point $\alpha_{c_2}$ at which the dimerized state of the sawtooth chain vanishes when $n$ increases. In the case of LSUB*n* approximation with $n \le 8$, although the CCM Néel solution denoted by $\theta_s = \pi$ （$\theta_c = 0$） always exists in the whole parameter region that we discuss, the energy of the Néel solution compares extremely poorly to that of ED in the large $\alpha$ parameter region and this is displayed in figure 11. Thus, we only use the spiral or the canted state solution of CCM to discuss the property of the sawtooth chain at that parameter region. Figure 11 shows the extrapolated results of CCM for $e$ using the scheme of equation (8) with the data set *n*={4, 6, 8}. It is obvious that the results of CCM agree well with those of ED.

To explain the above phenomenon, we also calculate the correlation function of the sawtooth chain by using DMRG under OBC [46]. Corresponding to the two sublattice structure of the sawtooth chain, two kinds of two-spin correlation functions are defined

$$
\begin{aligned}
C_{N/2,2i}(|N/2-2i|) &= \langle\psi|S_{N/2}\cdot S_{2i}|\psi\rangle \\
C_{N/2-1,2i-1}(|(N/2-1)-(2i-1)|) &= \langle\psi|S_{N/2-1}\cdot S_{2i-1}|\psi\rangle
\end{aligned}
\tag{14}
$$

where $|\psi\rangle$ is the ground state of the sawtooth chain. Figure 12 displays a logarithmic plot of the above two correlation functions [47]. We can see from figure 12 that the two correlation functions show an incommensurate behaviour, and the period of oscillation of $C_{N/2,2i}$ equals to that of $C_{N/2-1,2i-1}$ in the intermediate parameter region $1 < \alpha < 1.5$. This result suggests that the ground state of the sawtooth chain is an incommensurate spiral state as shown in figure 1(b) at that parameter region. In the large parameter region $\alpha > 1.5$, the character of $C_{N/2,2i}$ is different from that of $C_{N/2-1,2i-1}$. Thus, incommensurate spiral state is absent when $\alpha > 1.5$. According to the behaviour of the correlation function, one can conclude that it is more suitable to choose the spiral state than the canted state as the CCM's model state in the intermediate parameter region.

In order to investigate whether the sawtooth chain possesses the "true" canted order in the parameter region $\alpha > 1.5$, we calculate the magnetic order parameter *M* by using CCM. The results disclose that the value of *M* extrapolated to $n \to \infty$ using equation (8) is negative when $\alpha > 1.35$. For example, figure 13 displays the illustration of the extrapolation of the CCM LSUB*n* data for *M* when $\alpha = 1.55$ and $\alpha = 1.75$. Note that the abscissa of figure 13 is scaled according to the leading exponent of equation (8). One can observe that the extrapolation scheme of equation (8) fits the LSUB*n* data points well and $M < 0$ in the limit $n \to \infty$. Hence, the canted order is absent in the sawtooth chain. As the model still possesses short-range order in the large parameter region, we can call the ground state of the sawtooth chain quasi-canted state at that parameter region.

Besides the property of the ground state of the sawtooth chain, the results of CCM can also provide us the information of the nature of quantum phase transition occurring at $\alpha_{c_2}$. In the case of LSUB6 approximation, the canted solution for the ground state energy per site $e$ as a function of $\theta_c$ is shown in figure 9. As shown in figure 9, the curve has only one minimum when $\alpha < \alpha_{c_2}$. In the large parameter region $\alpha \geq \alpha_{c_2}$, besides the Néel solution, another minimum appears at a finite value of $\theta_c$. The appearance of the two-minimum structure for the ground state energy as a function of $\theta_c$ indicates that the transition from the dimerized state to the quasi-canted state belongs to a first-order phase transition [18]. The cross point of the ground state energy given by CCM in figure 10 may be another hint of the existence of the first-order phase transition [48].

To check the type of the transition occurring at $\alpha_{c_2}$ further, we also calculate the first derivative of the ground state energy of a finite system $dE_g / d\alpha$ by using DMRG under OBC and plot it in figure 14. As is apparent in that figure, $dE_g / d\alpha$, all of the results for various lattice sizes, *N*, display a discontinuity near $\alpha = 1.5$. Moreover, the height of the jump increases slightly when the system size *N* increases. Then, one can infer reasonably that the first derivative of the ground state energy is discontinuous in the thermodynamic limit. This result means, just as the CCM predicts, that there is a level crossing in the ground state of the sawtooth chain at the critical point $\alpha_{c_2}$. The numerical DMRG results provide us with a confirmation that the quantum phase transition occurring at $\alpha_{c_2}$ belongs to a first-order transition.

The critical point $\alpha_{c_2}$ drawn from CCM based on canted state is shown in table 1. A linear extrapolation [18], $\alpha_{c_2}^{\infty} = a_0 + a_1 n^{-1}$, gives that the estimate of the critical point $\alpha_{c_2}$ in the limit $n \to \infty$ is $\alpha_{c_2} = 1.42$. It is closed to the value of the critical point at which the dimerized state of the sawtooth chain vanishes given by ED.

## 4. Conclusions

In this paper, we studied the properties of the sawtooth chain by using CCM, ED and DMRG. The results of CCM based on Néel state show up in the following two points:

(1) Only CCM symmetry solution exists in the region $\alpha < \alpha_{c_1}$, which means that the ground state still has the translational lattice symmetry and the quasi-Néel state is always the ground state of the sawtooth chain at that parameter region.

(2) At the critical point $\alpha_{c_1}$, a non-symmetry solution also appears besides the symmetry solution. And the non-symmetry solution persists up to a termination point $\alpha_t$. Moreover, the exact dimerized state of the sawtooth chain at $\alpha = 1$ can be reproduced by using CCM symmetry-broken solution and the CCM symmetry-broken solution provides far better results than those of the symmetry solution when $\alpha_{c_1} < \alpha < \alpha_t$. Thus, the ground state of the sawtooth chain is in the dimerized state in the intermediate parameter region.

The results of ED indicates that the gapful dimerized state does exist in a parameter region $\alpha_{c_1} < \alpha < \alpha_{c_2}$. The critical point $\alpha_{c_2}$ can precisely be determined by using the results of the spin gap calculated by ED. To obtain the critical point $\alpha_{c_1}$, we studied the critical behavior of the spin stiffness fidelity susceptibility $\chi_\rho$ in the vicinity of $\alpha_{c_1}$. It is found that, similar to one-dimensional spin-1/2 $J_1 - J_2$ chain, $\alpha_{c_1}$ can be given accurately by analysing the behavior of the spin stiffness fidelity susceptibility in the vicinity of that critical point.

Although CCM based on Néel state is out of use in the strong $\alpha$ parameter region, we found that CCM based on spiral state and canted state can respectively provide good results for the ground state energy in the intermediate parameter region and the large $\alpha$ parameter region.

Numerical DMRG results indicate that the ground state is an incommensurate spiral state in the intermediate parameter region. We call the ground state of the sawtooth chain quasi-canted state because the “true” canted order is absent when $\alpha > \alpha_{c_2}$. The results of CCM and DMRG both indicate that the transition from the dimerized state to the quasi-canted state belongs to the first-order transition.

Combining the analysis of CCM with those of ED and DMRG, we can conclude that the overall phase diagram of the sawtooth chain is divided into three phases: quasi-Néel phase, dimerized phase with or without incommensurate spiral spin correlations and quasi-canted phase.

**Acknowledgments**

It is a pleasure to thank Dr. Damain Farnell for helpfully discussing with us and kindly providing us with his computation results to check the accuracy of our CCM programme. This work is supported by the National Natural Science Foundation of China (No. 10804053), the National Natural Science Foundation of China (No. 11104159), the Natural Science Foundation of Jiangsu Province under Grant Nos. BK20131428, the Natural Science Foundation of the Jiangsu Higher Education Institutions under Grant Nos. 13KJD140003, Qing Lan Project of Jiangsu Province, and the Scientific Research Foundation of Nanjing University of Posts and Telecommunications (No. NY211008).

**Table 1.** CCM results for the critical points $\alpha_{c_1}$ , $\alpha_c$ , and $\alpha_{c_2}$ .

| | $\alpha_{c_1}$ | $\alpha_c$ | $\alpha_{c_2}$ |
|---|---|---|---|
| LSUB4 | 0.949 | 1.16 | 1.25 |
| LSUB6 | 0.931 | 1.29 | 1.27 |
| LSUB8 | 0.920 | 1.33 | 1.35 |
| LSUB10 | 0.908 | — | — |
| LSUB12 | 0.894 | — | — |
| LSUB14 | 0.880 | — | — |

# Figure captions

Fig.1: The sketches of the classical canted state (a) and the spiral state (b) of the sawtooth chain. $\theta_c$ or $(2i-1)\theta_s$ ($(2i)\theta_s$) measures the deviation of the classical spins from the *z* axis.

Fig.2: The two-spin nearest-neighbor ket-state correlation coefficient at the LSUB14 level of approximation. The full line without symbols shows the symmetry solution ($S_2^a = S_2^b$). The correlation coefficient $S_2^a$ and $S_2^b$ for the non-symmetry solution are separately displayed by the open and filled circles. The termination point of the CCM equations is indicated by the boxes.

Fig.3: The ground state energy per site $e$ versus $\alpha$ using CCM based on Néel state at the LSUB14 level of approximation and ED.

Fig.4: The spin gap $\Delta$ versus $\alpha$ using CCM and ED with PBC. The inset shows the behavior of $\Delta$ given by ED when $\alpha > 1.2$.

Fig.5: The spin gap $\Delta$ versus $\alpha$ using ED with OBC.

Fig.6: The reduced fidelity susceptibility $\chi_\rho / N$ as a function of $\alpha$ for various size *N*.

Fig.7: Finite-size scaling of $\alpha_{\min}$ of $\chi_\rho / N$ versus $N^{-1}$. The solid line is the fit line.

Fig.8: The ground state energy per site $e$ versus the spiral angle $\theta_s$ using CCM based on spiral state at the LSUB6 level of approximation.

Fig.9: The ground state energy per site $e$ versus the canted angle $\theta_c$ using CCM based on canted state at the LSUB6 level of approximation.

Fig.10: The ground state energy per site $e$ is shown for each LSUB*n* approximation.

Fig.11: The ground state energy per site $e$ obtained from CCM and ED in the large $\alpha$ region.

Fig.12: The logarithmic plot of $(N/4-i)^{0.5} |< S_{N/2} S_{2i} >|$ and $(N/4-i)^{0.5} |< S_{N/2-1} S_{2i-1} >|$ obtained from DMRG for several values of $\alpha$ in a system with *N*=120.

Fig.13: Illustration of the extrapolation (solid lines) of the CCM LSUB*n* data (symbols) for the magnetic order parameter *M*.

Fig.14: The first derivative of the ground state energy $dE_g / d\alpha$ as a function of $\alpha$ obtained from DMRG for *N*=24, *N*=32, and *N*=40.

**Figure 1**

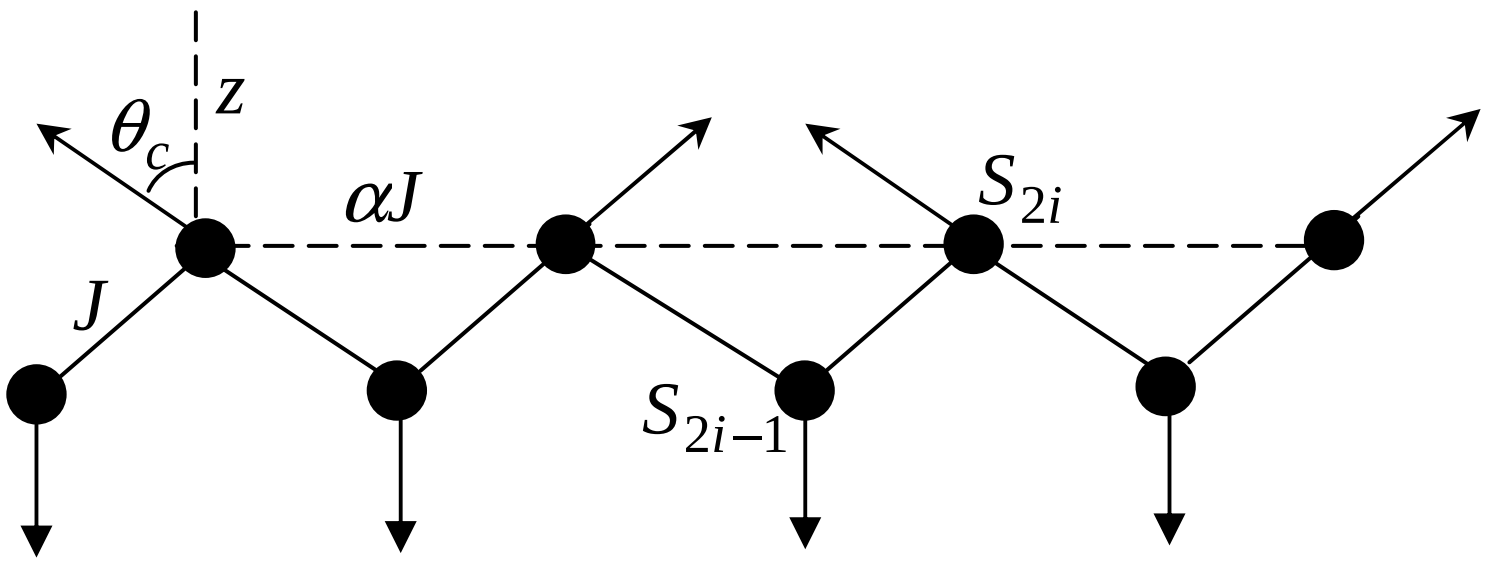

**(a)**

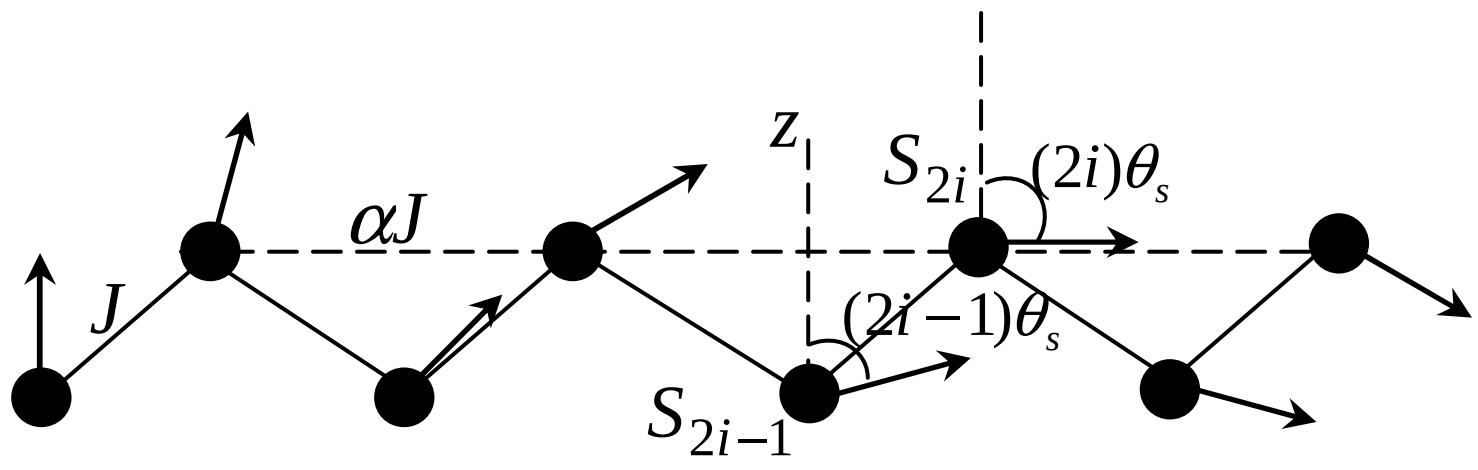

**(b)**

**Figure 2**

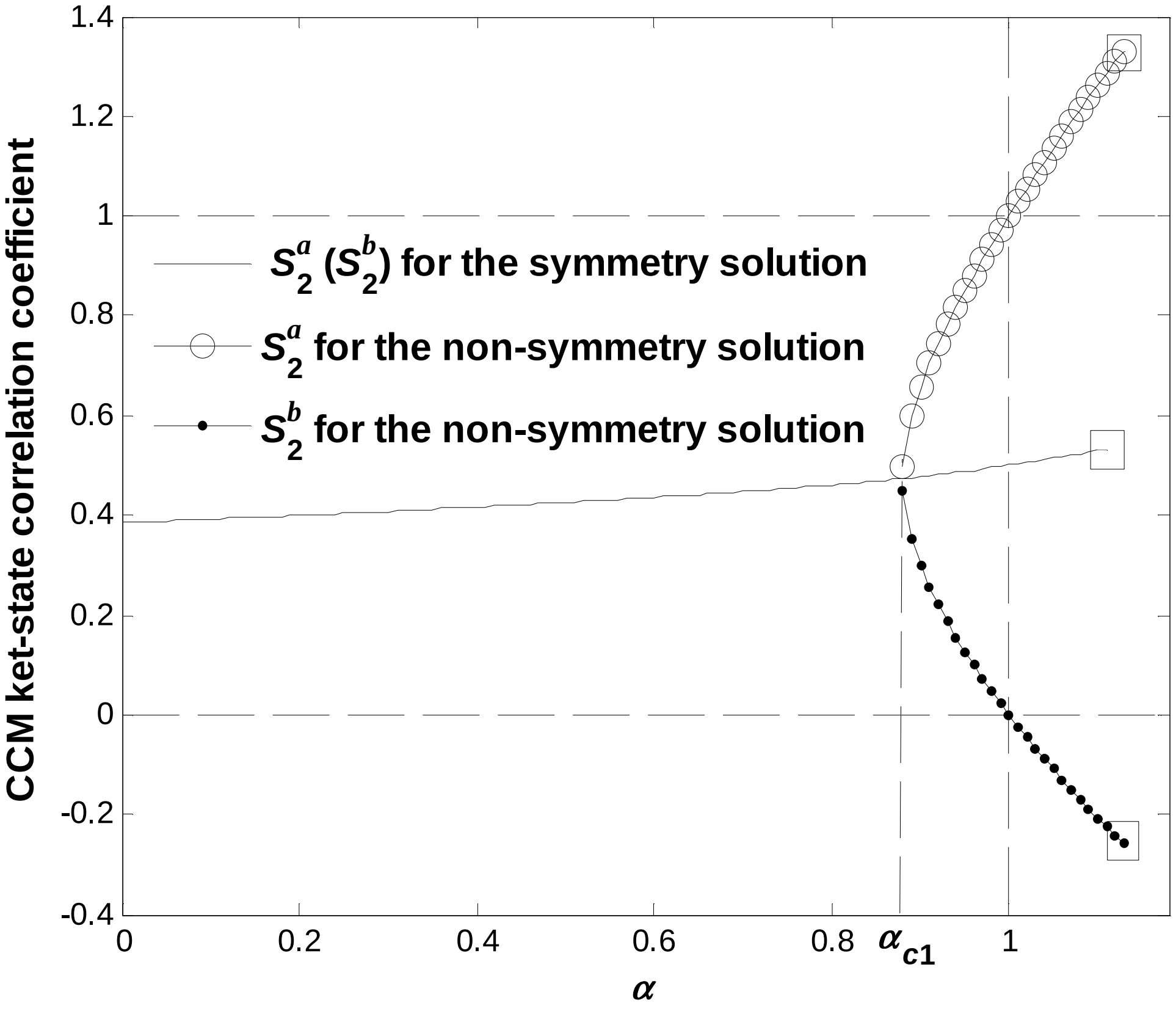

**Figure 3**

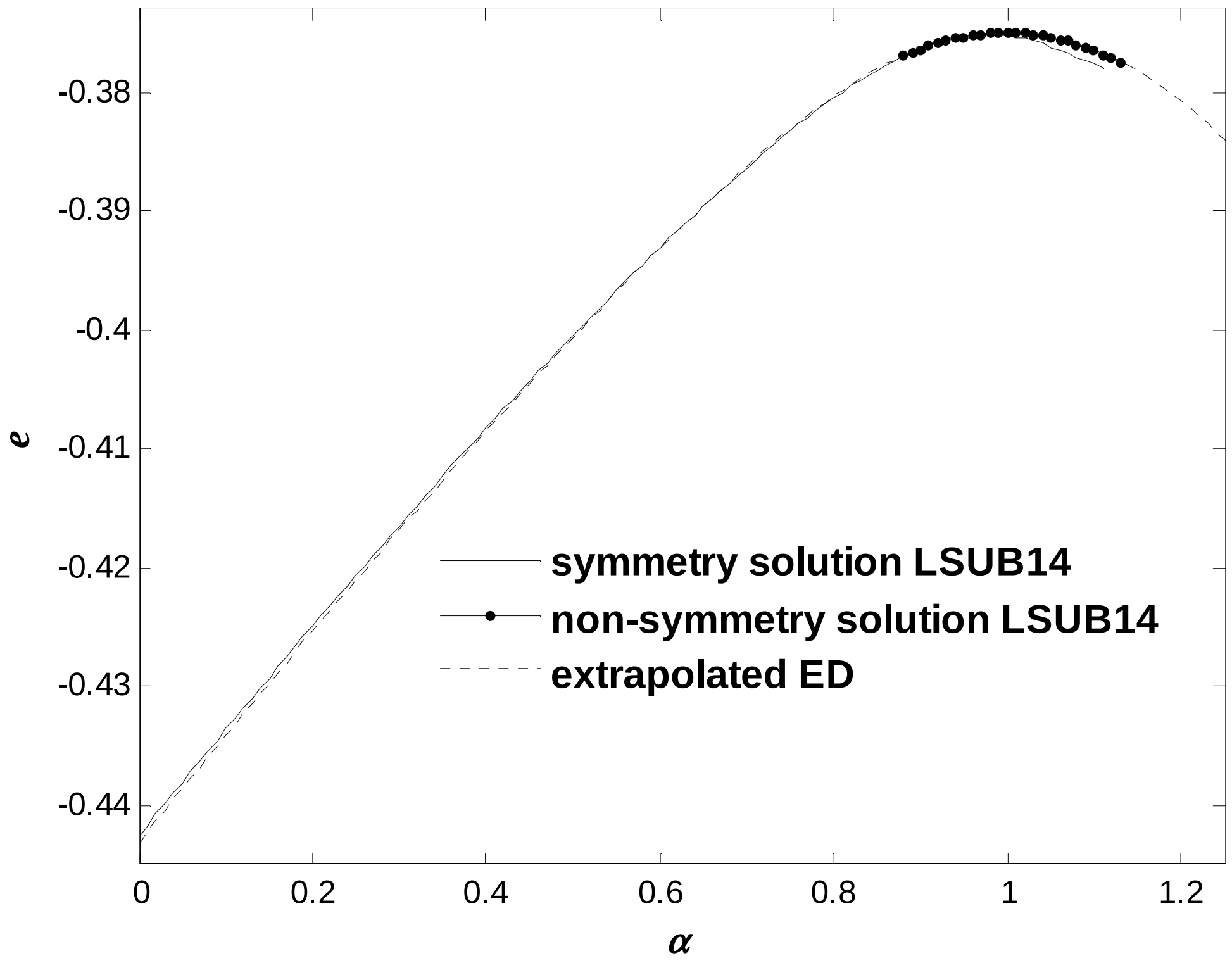

**Figure 4**

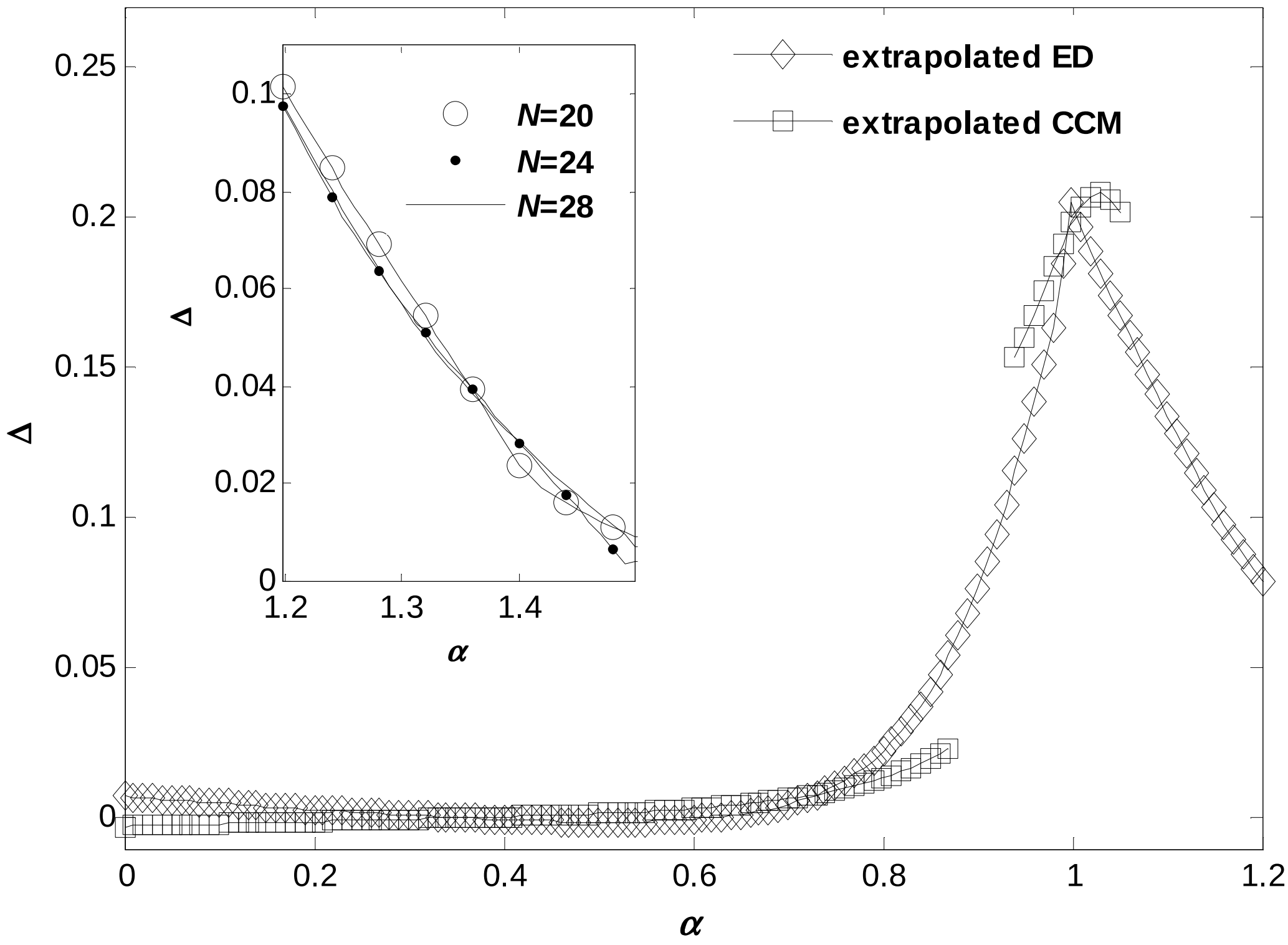

**Figure 5**

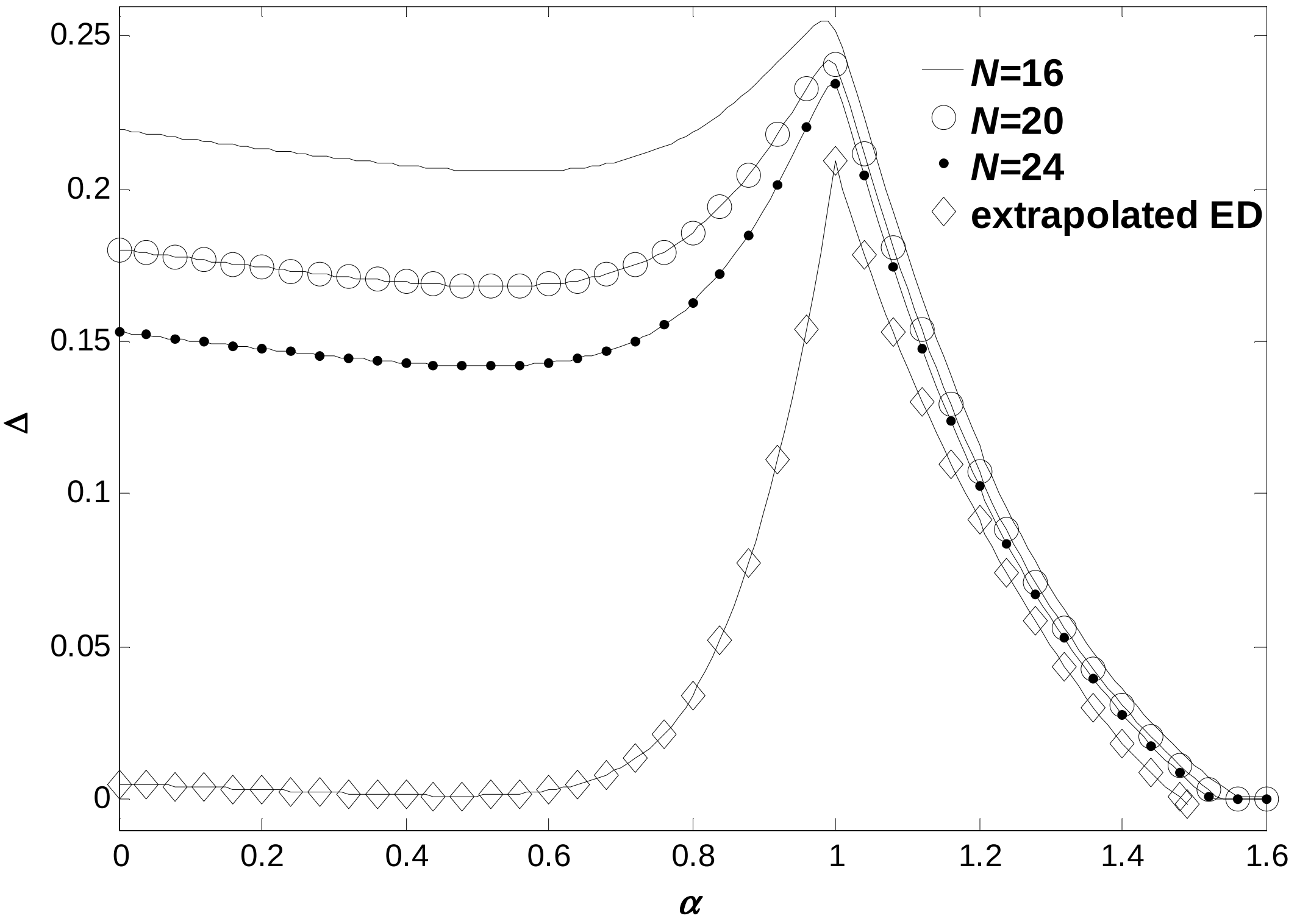

**Figure 6**

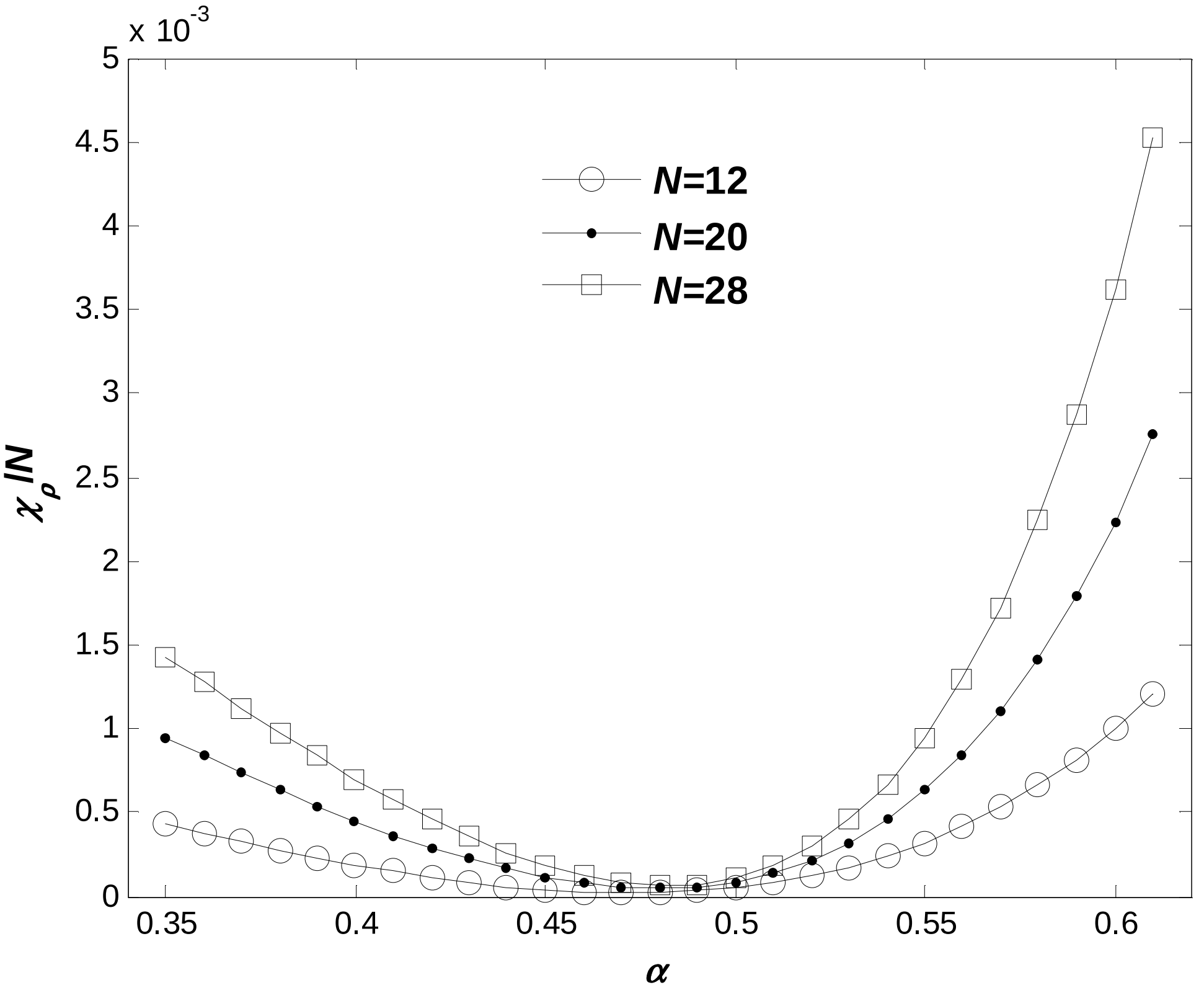

**Figure 7**

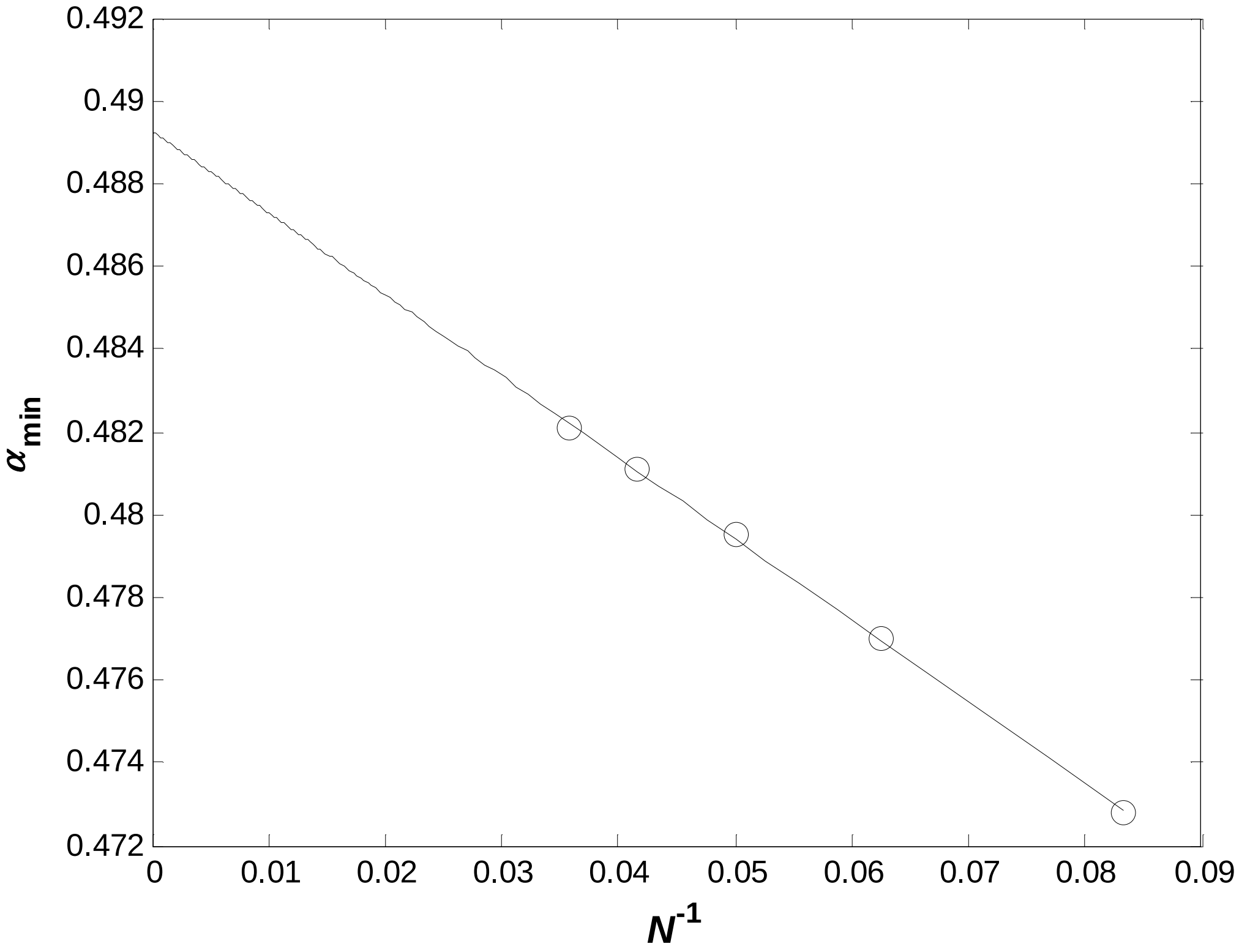

# Figure 8

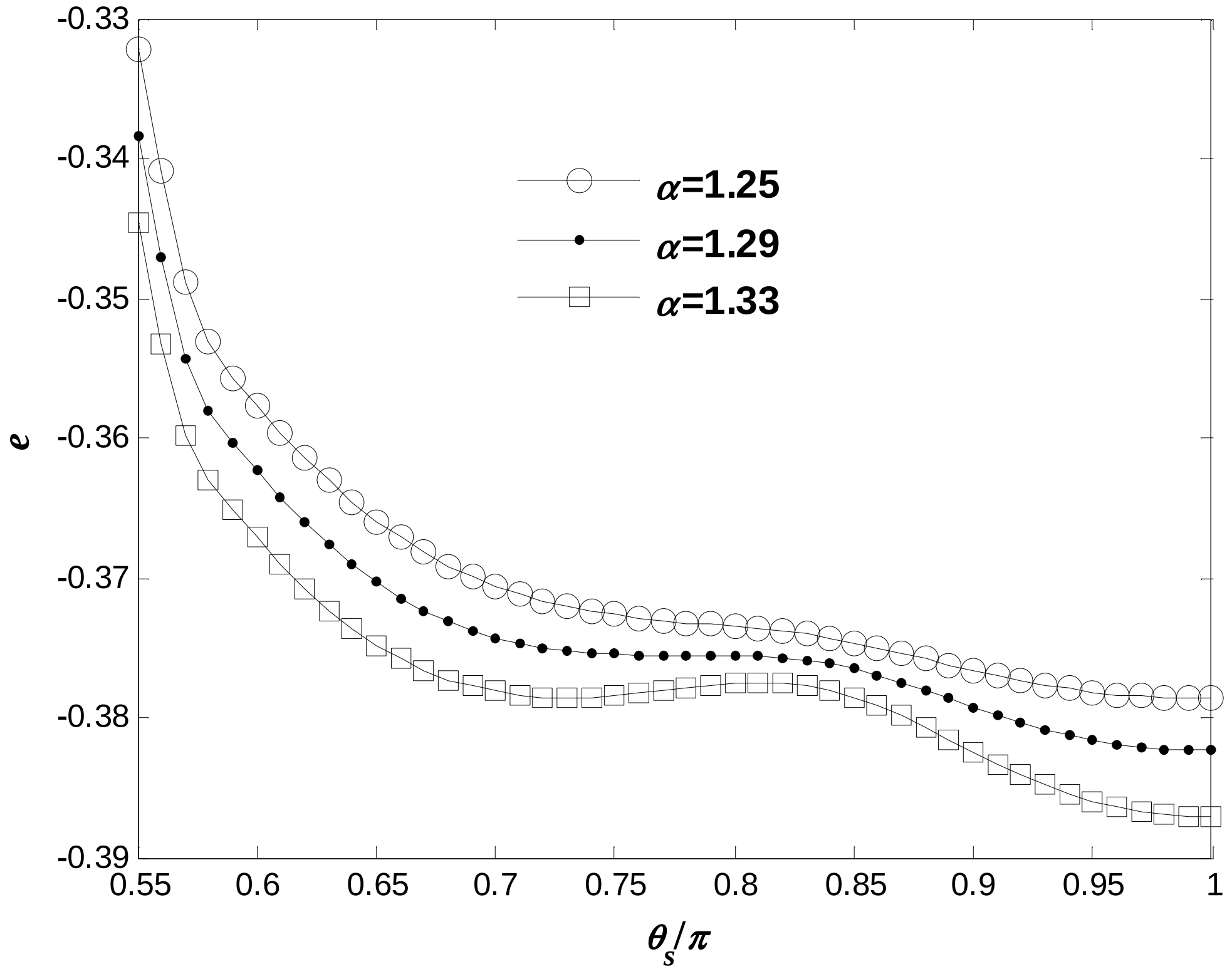

**Figure 9**

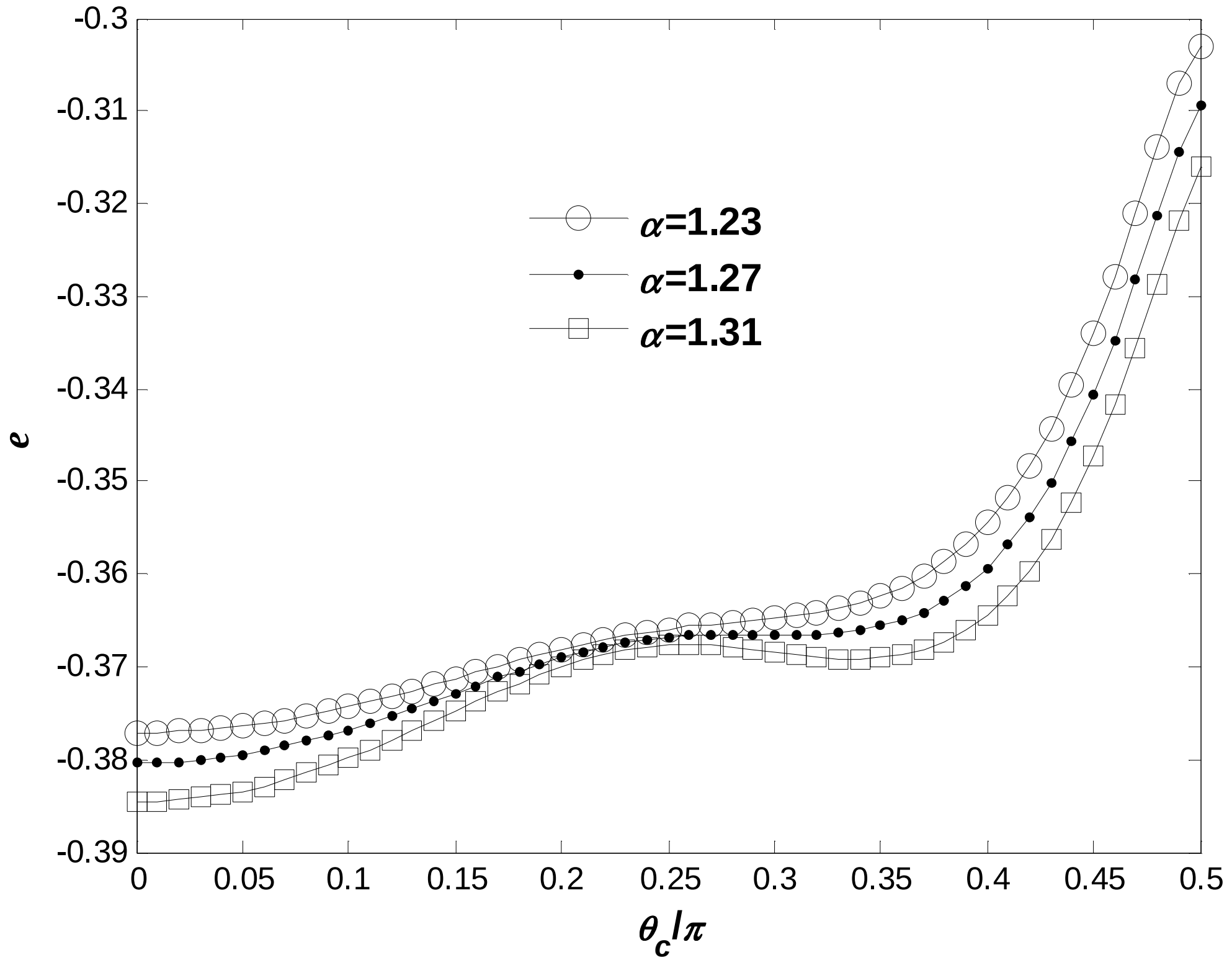

**Figure 10**

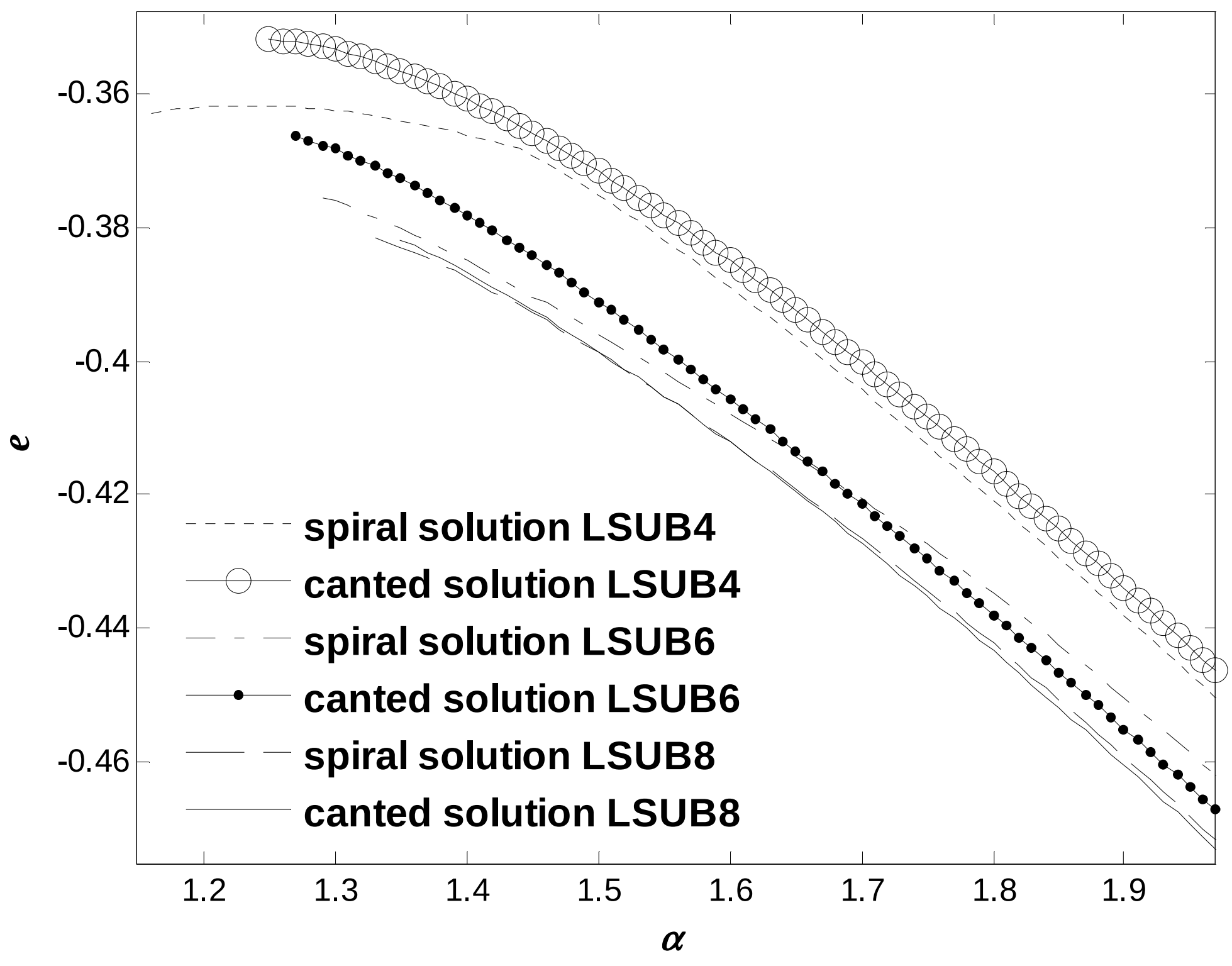

**Figure 11**

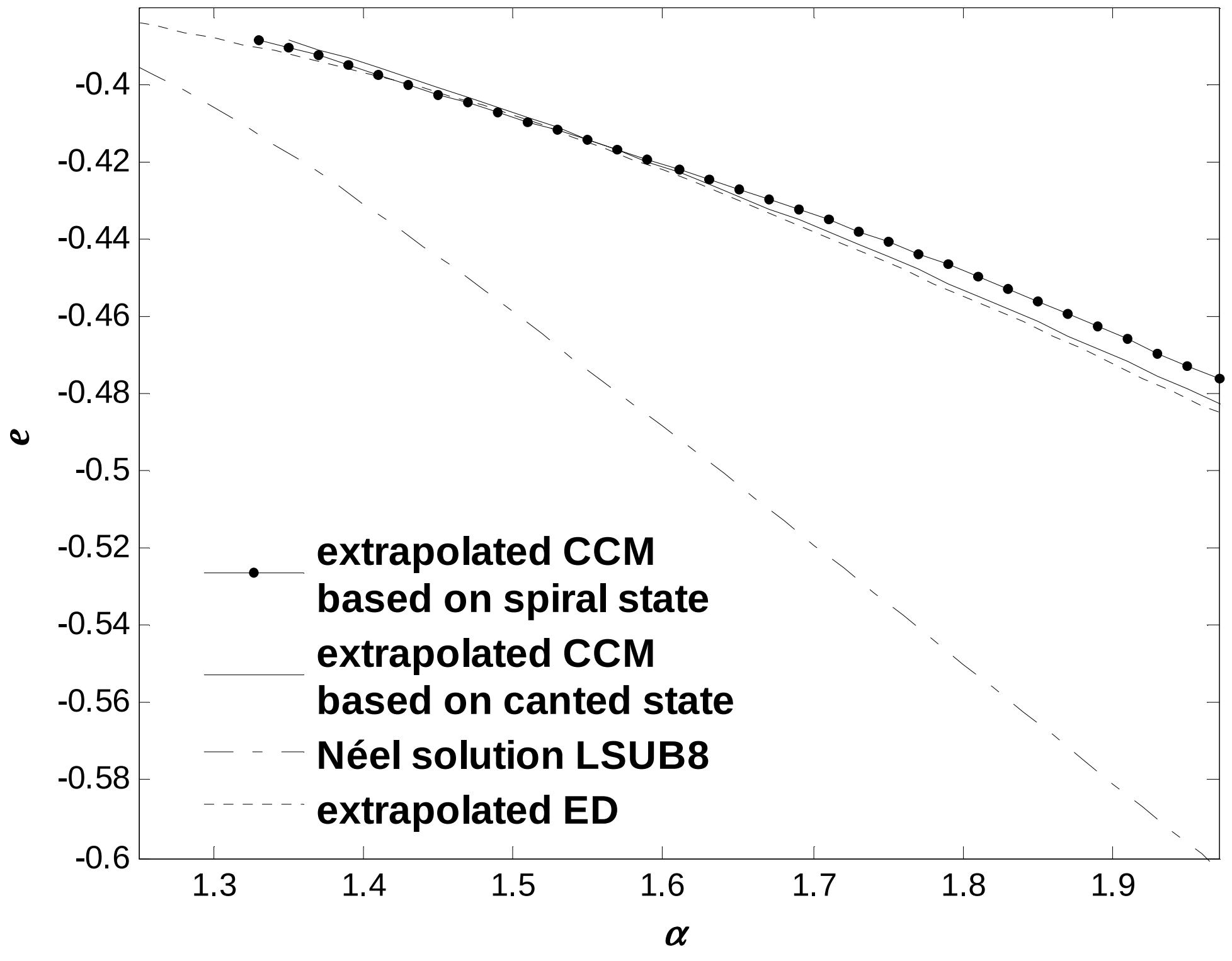

## Figure 12

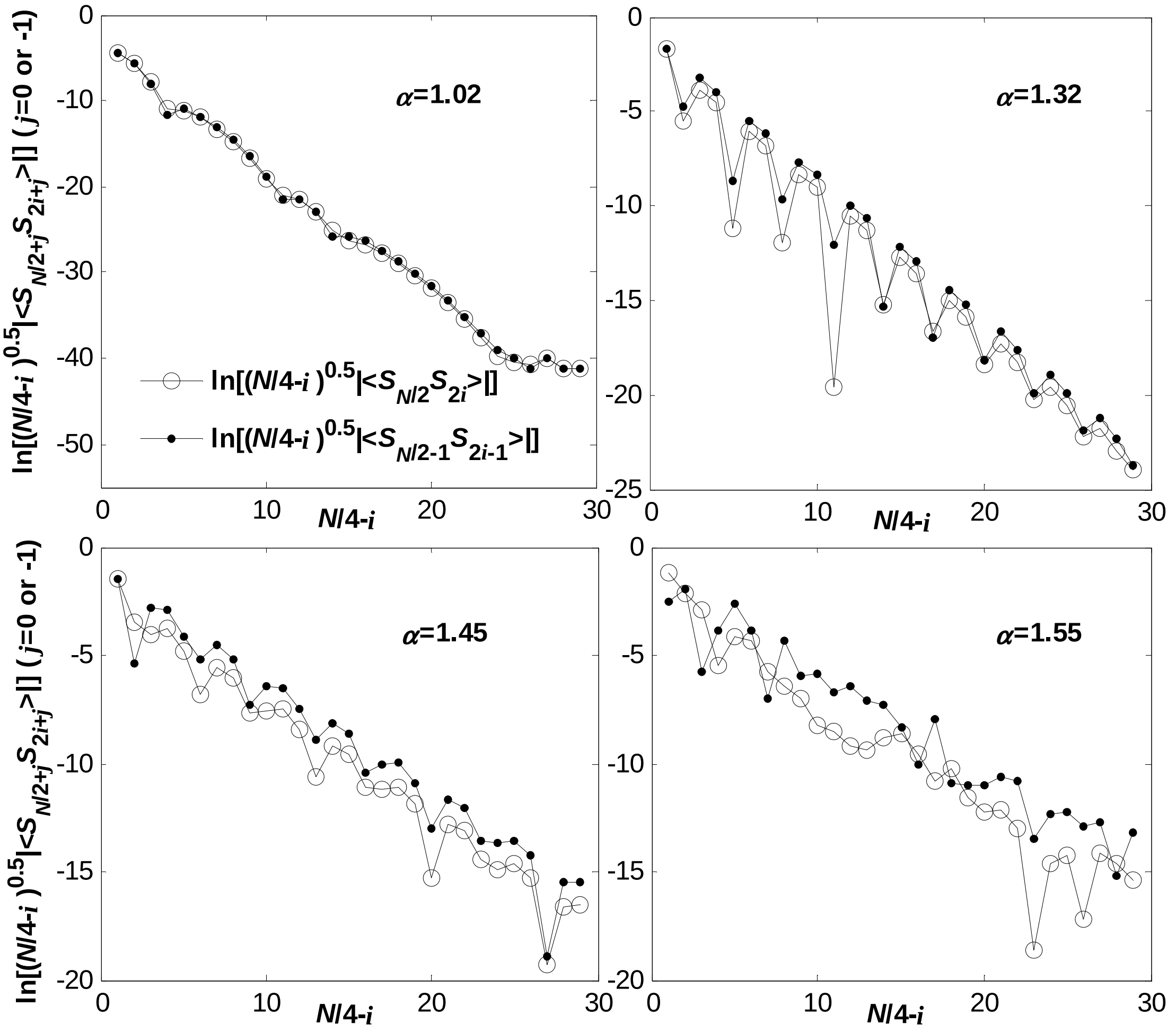

**Figure 13**

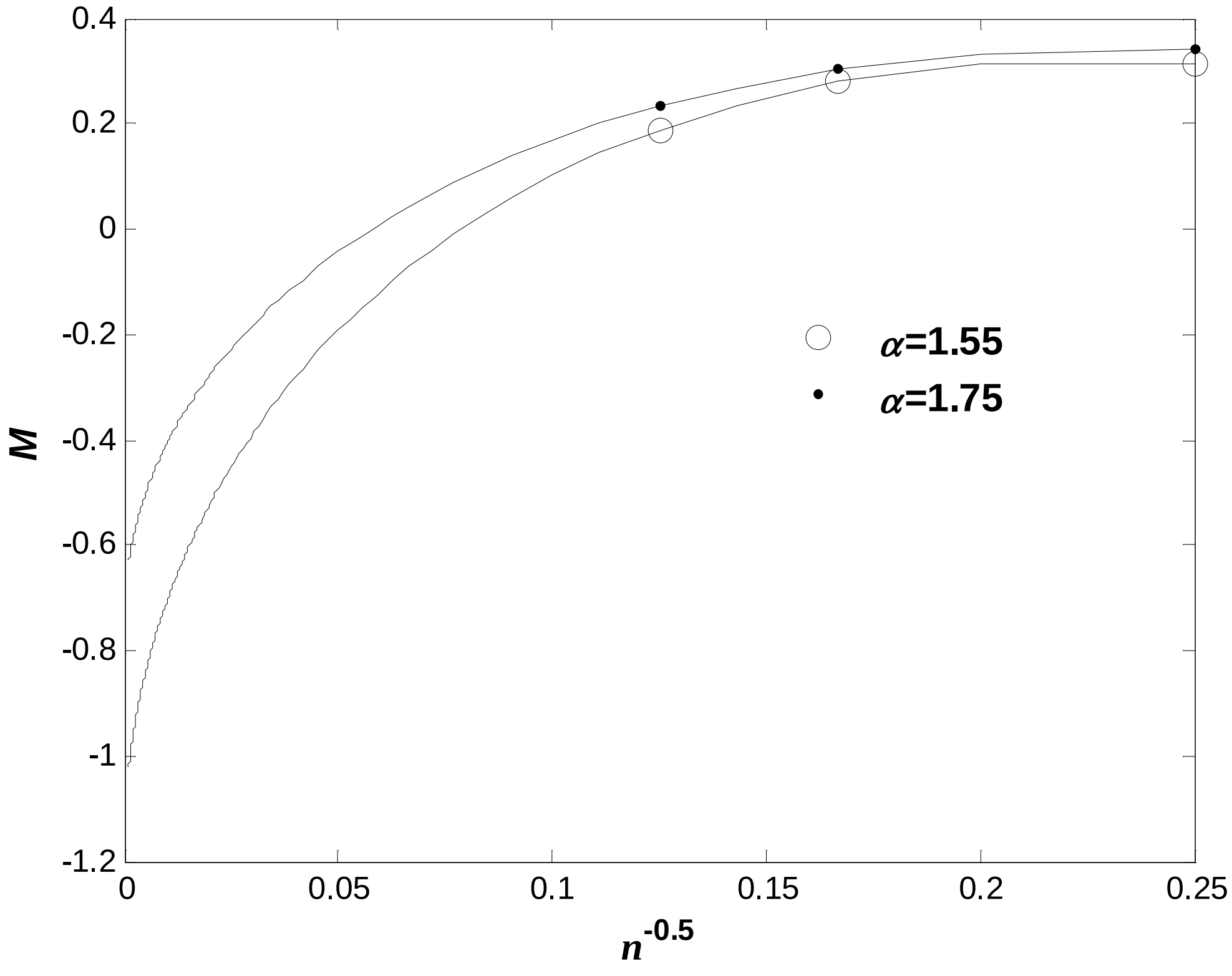

# Figure 14

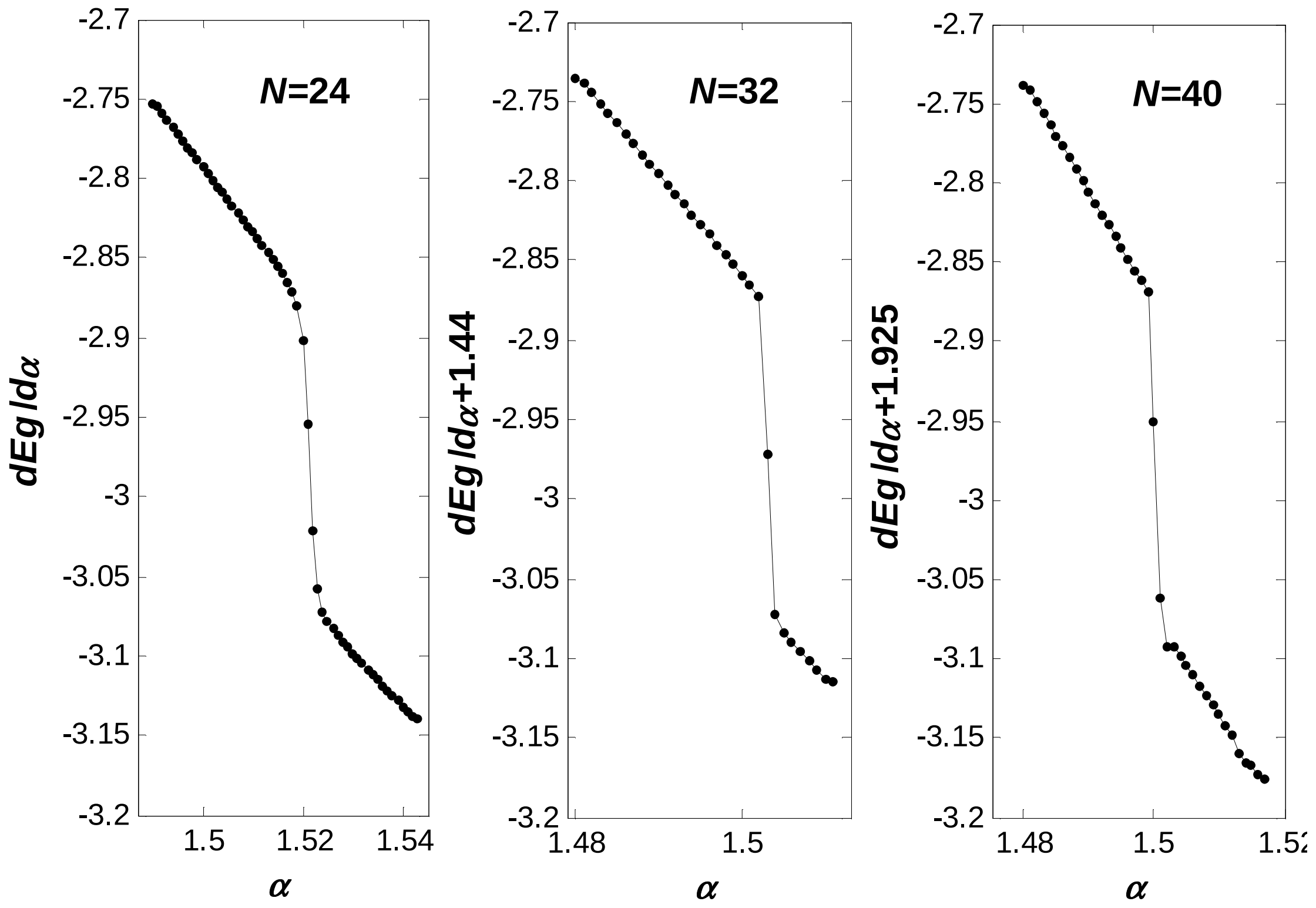